\documentclass{jpp}
\usepackage{graphicx,psfrag}
\usepackage{epstopdf, epsfig}
\usepackage{graphicx, amsfonts, amsmath, amssymb}

\newtheorem{definition}{Definition}

\shorttitle{Non-modal tearing instability}
\shortauthor{D. MacTaggart}

\title{The non-modal onset of the tearing instability}

\author{D. MacTaggart\aff{1}
  \corresp{\email{david.mactaggart@glasgow.ac.uk}}
}

\affiliation{\aff{1}School of Mathematics \& Statistics, University of Glasgow, UK, G12 8SQ}

\newcommand{\be}{\begin{equation}}
\newcommand{\en}{\end{equation}}
\def\pt{{\partial}}

\def\uv{{\boldsymbol u}}
\def\vv{{\boldsymbol v}}

\def\jv{{\boldsymbol j}}
\def\ev{{\boldsymbol e}}
\def\Bv{{\boldsymbol B}}

\def\cd{{D}}

\def\cl{{\mathsfbi{L}}}
\def\cm{\boldsymbol{\mathsfbi{M}}}
\def\dm{\boldsymbol{\mathsfbi{D}}}
\def\sm{\boldsymbol{\mathsfbi{S}}}
\def\wm{\boldsymbol{\mathsfbi{W}}}
\def\tm{\boldsymbol{\mathsfbi{T}}}
\def\rrm{\boldsymbol{\mathsfbi{R}}}
\def\am{\boldsymbol{\mathsfbi{A}}}
\def\bm{\boldsymbol{\mathsfbi{B}}}
\def\vm{\boldsymbol{\mathsfbi{V}}}
\def\qm{\boldsymbol{\mathsfbi{Q}}}

\def\se{{\sigma_{\epsilon}}}
\def\grad{\boldsymbol\nabla}

\begin{document}

\maketitle

\begin{abstract}
We investigate the onset of the classical magnetohydrodynamic (MHD) tearing instability (TI) and focus on non-modal (transient) growth rather than the tearing mode. With the help of pseudospectral theory, the operators of the linear equations are shown to be highly non-normal, resulting in the possibility of significant transient growth at the onset of the TI. {This possibility increases as the Lundquist number $S$ increases.}  In particular, we find evidence, numerically, that the maximum {possible} transient growth, {measured} in the $L_2$-norm, for the classical setup of {current sheets unstable to} the TI, scales as $O(S^{1/4})$ on time scales of $O(S^{1/4})$ {for $S\gg 1$}. This behaviour is much faster than the time scale $O(S^{1/2})$ when the solution behaviour is dominated by the tearing mode. {The size of transient growth obtained is dependent on the form of  the initial perturbation.} Optimal initial conditions for the {maximum possible} transient growth are determined, which take the form of wave packets and can be thought of as noise concentrated at the current sheet. We also examine how the structure of the eigenvalue spectrum relates to physical quantities.
\end{abstract}

\section{Introduction}
In magnetohydrodynamics (MHD), the tearing instability (TI) occurs in highly sheared magnetic field configurations called {current sheets}. In a current sheet there is a thin (compared to larger length scales outside the current sheet) layer of intense current density where the magnetic field changes direction rapidly. If the conditions of the TI are met, the current sheet begins to `tear' or, to be more precise, the topology of the magnetic field changes to form multiple islands (or plasmoids in three dimensions) of magnetic flux. Since the seminal work of \cite{furth63}, the onset of the TI has been traditionally studied using normal mode analysis, to the extent that the terms `tearing instability' and `tearing mode' are often used synonymously. 

{Recent studies that address the linear onset of TI in high aspect ratio current sheets, also known as the plasmoid instability (PI), \citep[e.g.][]{loureiro07, bhattacharjee09,pucci14,uzdensky16,tenerani16} do so from the point of view of normal mode analysis. For the TI (and the PI), however, normal mode analysis cannot give a complete picture of its linear onset.}  The operators in the equations describing the onset of the TI are non-normal. This means that eigenmodes are not orthogonal and for the application in hand are heavily ill-conditioned \citep{borba94}. Therefore, although eigenmodes may be damped as $t\rightarrow\infty$, they can result in significant transient (or algebraic) growth within a finite time. Performing normal mode analysis on equations with non-normal operators results in the translation to a later time when the transient growth has been damped away. {Therefore, if significant transient growth is possible, it is ignored in normal mode analysis}.  

Although stability theory in plasma physics is dominated by normal mode (eigenvalue) analysis, studies of non-modal behaviour are on the increase. In the MHD literature, one early suggestion that subcritical behaviour may be important for the tearing instability was made by \cite{dahlburg83}, although the mechanism was thought to be nonlinear rather than linear. Later, \cite{dahlburg94} studied the algebraic growth of current sheets in ideal MHD as a possible route to turbulent reconnection through the creation of smaller length scales. \cite{borba94} investigated the eigenmodes of resistive MHD using pseudospectra but did not focus on the TI. They argue that the non-orthogonality of the eigenmodes implies that normal mode analysis can only describe instability growth on a long time scale (of $O(S^{1/2})$, where $S$ is the Lundquist number that will be defined later). Other researchers have recognized the importance of non-modal growth in other MHD applications, including kinematic dynamo theory \citep[e.g.][]{farrell991,farrell992,livermore06,chen18}, the magnetorotational instability \citep[e.g.][]{squire14a,squire14b} and the tearing instability \citep{dmac17}. There is also a growing interest in the subcritical transition to turbulence in tokamak plasmas \citep[e.g.][]{landremann15,vw16} and the non-modal consequences of shearing on microinstabilities \citep[e.g.][]{newton10}.

The purpose of this article is to investigate the non-modal transient growth at the onset of the classical TI. In particular, our aim is to determine the dependence of the {maximum possible} transient growth on the Lundquist number and to understand the relationship between the eigenvalue spectrum and the underlying physics. We solve the linearized equations numerically and use pseudospectral theory to help us understand how the spectrum relates to (a) the transient growth and (b) the optimal initial conditions that give rise to the maximum possible transient growth.

\section{Model description}

To study the TI, we consider the non-dimensional, incompressible, visco-resistive MHD equations

\be\label{mhd1}
\frac{\pt\uv}{\pt t}+(\uv\cdot\grad)\uv = - \grad p+(\grad\times\Bv)\times\Bv +\frac{1}{Re}\nabla^2\uv,
\en

\be\label{mhd2}
\frac{\pt\Bv}{\pt t} = \grad\times(\uv\times\Bv)+\frac{1}{S}\grad^2\Bv,
\en

\be\label{mhd3}
\grad\cdot\Bv = \grad\cdot\uv = 0,
\en
where $\Bv$ is the magnetic field, $\uv$ is the velocity, $p$ is the plasma pressure, $Re$ is the Reynolds number and $S$ is the Lundquist number. 

For our background equilibrium,
\be\label{back}
\quad p_0 = p_0(x), \quad {\Bv}_0 = B_0(x)\ev_z, \quad {\uv}_0 = U_0(x)\ev_z,
\en
where the subscript 0 corresponds to the equilibrium and
\begin{eqnarray}
 \quad {\bf 0} &=& -\grad p_0 + (\grad\times\Bv_0)\times\Bv_0 + \frac{1}{Re}\grad^2\uv_0,\label{equil1}\\
 \quad {\bf 0} &=&\grad\times(\uv_0\times\Bv_0).\label{equil2}
\end{eqnarray}
Magnetic diffusion is not included in the background equilibrium (\ref{equil2}) as we are only interested in time scales much shorter than the global magnetic diffusion time scale. Clearly, for the assumed forms of the equilibrium magnetic and velocity fields (\ref{back})$_2$ and (\ref{back})$_3$, equation (\ref{equil2}) is satisfied. Once $\uv_0$ and $\Bv_0$ are chosen, the background pressure $p_0$ is determined from equation (\ref{equil1}).

We will now linearize equations (\ref{mhd1}) and (\ref{mhd2}) about a background equilibrium by setting $(\uv,\Bv,p)=(\uv_0,\Bv_0,p_0)+(\uv_1,\Bv_1,p_1)$ and focus on the two-dimesional version of the equations. Assuming perturbations of the form
\be
\uv_1 = [u(x,t),0,u_z(x,t)]^T\exp(ikz), \quad \Bv_1 = [b(x,t),0,b_z(x,t)]^T\exp(ikz),
\en
the linearized form of equations (\ref{mhd1}) and (\ref{mhd2}) can be written as
\begin{eqnarray}
\frac{\pt}{\pt t}(\cd^2-k^2)u &=& {L}_{B_0}b - {L}_{U_0}u  + \frac{1}{Re}(D^2-k^2)^2u,\label{main1} \\
\frac{\pt b}{\pt t} &=&  ik(B_0u+U_0b) +\frac{1}{S}(\cd^2-k^2)b, \label{main2}
\end{eqnarray}
where 
\be
L_{U_0} = ik[U_0(\cd^2-k^2)-U_0''],\quad  {L}_{B_0} = ik[B_0(\cd^2-k^2)-B_0''], \quad \cd=\pt/\pt x,
\en 
and the prime refers to differentiation with respect to $x$ in the background equilibrium fields. {Equations (\ref{main1}) and (\ref{main2}) have essentially the same form as those obtained with reduced MHD in the presence of a large guide field \citep[e.g.][]{loureiro07}.}

To complete the setup of the model, we require boundary conditions for equations (\ref{main1}) and (\ref{main2}). In this paper we will consider {no-slip and perfectly conducting} boundary conditions,
\be\label{bc}
u=\cd u=b=0\quad {\rm at}\quad x=\pm d,
\en
where $d$ is a non-dimensional distance. Since the tearing instability grows in a thin boundary layer at $x=0$, the choice of boundary conditions should not have a large effect on the initial development of the instability if $d$ is sufficiently large.

To facilitate the discussion of our analysis later, we rewrite equations (\ref{main1}) and (\ref{main2}) in the form

\be\label{ivp}
\frac{\pt}{\pt t}M\vv = L\vv,
\en
where $\vv=(u,b)^{\rm T}$,

\be\label{mm}
M=\left(\begin{array}{cc}
\cd^2-k^2 & 0\\
0&{I}	
\end{array}\right), \quad L = \left(\begin{array}{cc}
\frac{1}{Re}(\cd^2-k^2)^2-L_{U_0} & L_{B_0}\\
ikB_0&ikU_0+\frac{1}{S}(\cd^2-k^2)
\end{array}\right),
\en
and
${I}$ represents the identity operator.

\section{Numerical implementation}
In this section we describe briefly the numerical techniques used to discretize and study the behaviour of equation (\ref{ivp}). Henceforth, we will describe matrices rather than operators and eigenvectors rather than eigenmodes as the equations will be discretized. To signify this change, the notation for a matrix will have the same letter as the operator it represents but it will now be in bold, e.g. $A$ is an operator and $\mathsfbi{A}$ is the finite matrix discretization of that operator. In order not to introduce too much extra notation, eigenvectors will have the same notation as the eigenmodes.  

\subsection{Solving the eigenvalue problem}
Assuming a time dependence of $\exp(\sigma t)$, the initial value problem (\ref{ivp}) becomes the generalized eigenvalue problem
\be\label{eig_new}
\sigma\cm\vv = \cl\vv.
\en
The matrix $\cm^{-1}\cl$ involves (discrete) derivates (up to fourth order) in $x\in[-d,d]$. It is trivial to move from this domain to $[-1,1]$ and so we discretize the equations using a Chebyshev pseudospectral method \citep{t99,t00}. If $N$ be a positive integer, the $N+1$ Chebyshev points are given by
\be
x_i = \cos\left(\frac{i\pi}{N}\right), \quad i = 0,\dots,N.
\en
Note that we could define the $x_i$ with a minus sign in front so as to move from -1 to 1. However, due to the symmetry of our problem, there is no need to make this step. On the domain $[-1,1]$ the first order spectral differentiation matrix ${\dm}_N$ is given by
\[
({\dm}_N)_{00} = \frac{2N^2+1}{6}, \quad ({\dm}_N)_{NN} = -\frac{2N^2+1}{6},
\]
\be
({\dm}_N)_{jj} = -\frac{x_j}{2(1-x_j^2)} \quad {\rm for} \quad 1\le j\le N-1,
\en
\[
({\dm}_N)_{ij}= \frac{c_i}{c_j}\frac{(-1)^{i+j}}{x_i-x_j} \quad {\rm for} \quad i\ne j.
\]
The above coefficients are defined as
\be
c_i=\left\{\begin{array}{ccc}
2 & {\rm for}& i = 0\,\, {\rm or}\,\, N, \\
1 & {\rm for}&  1\le i\le N-1.\end{array}\right.
\en
The second order differentiation matrix is given simply by ${\dm}_N^2$. Due to the boundary conditions (\ref{bc}), we strip the first and last rows of ${\dm}_N$ and ${\dm}_N^2$.

The above matrices are constructed via polynomial interpolation \citep{t00}.   To define a fourth order differentiation matrix, we require a polynomial interpolant that satisfies two more boundary conditions than that for the second order differentiation matrix. Again following \cite{t00}, the resulting fourth order differentiation matrix is 
\be
{\sm}_N = [{\rm diag}(1-x_j^2){\dm}_N^4-8{\rm diag}(x_j){\dm}_N^3-12{\dm}_N^2]{\rm diag}\left(\frac{1}{1-x_j^2}\right),
\en
where $j=1,\dots,N-1$ after stripping the first and last rows.

With all the differentiation matrices defined, the matrices of the eigenvalue problem are constructed as they are displayed in (\ref{mm}). We then solve the eigenvalue problem (\ref{eig_new}) using the QZ algorithm \citep{golub96}.

\subsection{Quadrature}\label{quad}
Once the eigenvalue spectrum is obtained, this describes the asymptotic phase of the linear onset of the TI. To aid our understanding of how the transient growth relates to the spectrum, we require a generalization of the eigenvalue spectrum known as the pseudospectrum \citep{borba94,t05}. The calculation of pseudospectra (described in more detail later) and other useful quantities will require the evaluation of norms. To evaluate norms accurately, we must take into account appropriate quadrature weights for the spectral discretizations on an irregular grid. Following the description given in \cite{t99}, we consider a weight matrix of the form
\be
\wm={\rm diag}(w_1,\dots,w_N,w_1,\dots,w_N),
\en   
with Gauss-Lobatto weights defined by
\be
w_j^2=\frac{\pi\sqrt{d^2-x_j^2}}{2(N+1)}.
\en
If $\|\cdot\|$ denotes the weighted vector norm that approximates the continuous $L_2$-norm, then
\be\label{vnorm}
\|\uv\| = \|\wm\uv\|_2,
\en
for some vector $\uv$ of length $2N$. The corresponding matrix norm is
\be\label{mnorm}
\|\am\| = \|\wm\am\wm^{-1}\|_2,
\en
where $\am$ is a $2N\times 2N$ matrix. More detailed descriptions of these and related results can be found in \cite{t99,t05,reddy93}. Throughout the rest of this article, the vector and matrix norms that we will use are those defined in (\ref{vnorm}) and (\ref{mnorm}).
\subsection{Matrix projection}
One practical issue related to the numerical solution of the eigenvalue problem is that `spurious eigenvalues' \citep[e.g.][]{bourne03} are generated by the numerical scheme which do not have any physical interpretation related to the TI. There are various methods of removing these {values}, such as solving the equivalent adjoint eigenvalue problem and removing any eigenvalues that do not appear in both the original and adjoint calculations \citep[e.g.][]{stewart09}. In this article, we bypass the problem of spurious eigenvalues by projecting $\cm^{-1}\cl$ onto a lower-dimensional subspace. That is, we focus on a (physically interesting) part of the complex plane and only consider the eigenvalues in this region, {cutting out the spurious eigenvalues. } This approach also aids to accelerate calculations. There are many ways to achieve projection. In this article we make use of the QR algorithm \citep{golub96}. Let $\bm=\wm\cm^{-1}\cl\wm^{-1}$ and $\vm$ be an $2N\times n$ matrix whose columns are selected linearly independent eigenvectors of $\bm$ satisfying $\bm\vm=\vm\dm$ for some $n\times n$ diagonal matrix $\dm$ of corresponding eigenvalues. If $\vm=\qm\rrm$ is a QR decomposition of $\vm$, we can find an upper-triangular $n\times n$ matrix
\be
\tm = \rrm\dm\rrm^{-1},
\en 
which is the matrix representation of the projection of $\bm$ onto a space of the selected eigenvectors.

{Later, we will investigate what parts of the eigenvalue spectrum are `physically interesting', i.e. what parts contribute the largest transient growth.  Knowing these locations will allow us to select a suitable projection using the above method.}

\section{Transient growth analysis}
\subsection{Parameter selection}
We will consider a domain size given by $d=10$, which allows us to make comparisons with previous work \citep{dmac17} and is a value large enough to not allow the boundary conditions to interfere strongly with the onset of the TI. Here we will only focus on the classical TI and set $U_0=0$ and $Re=10^6$. In future work we will consider the effects of different background flows and current sheet thicknesses. The background magnetic field equilibrium is given by
\be
B_0(x) = \tanh(x),
\en
corresponding to the Harris sheet \citep{harris62}. 
\subsection{Spectra}
\begin{figure}

	\centering
	
		{\includegraphics[scale=0.5,keepaspectratio]{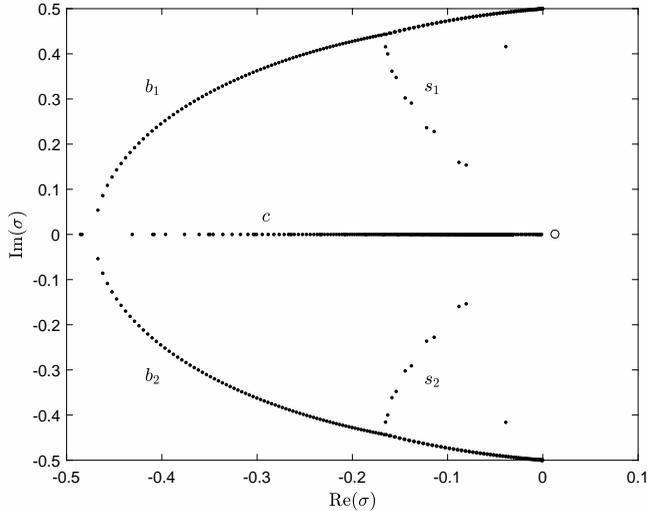}}
	\caption{Eigenvalue spectrum with $\Real(\sigma)\ge-0.5$ for $S=10^3$ and $k=0.5$. Eigenvalues are shown as solid dots apart from the unique eigenvalue corresponding to the tearing mode, which is shown as a hollow circle.}
\label{spectrum}
\end{figure}

The eigenvalue spectrum for the onset of the TI consists of a branched structure below $\Real(\sigma)=0$ and a unique eigenvalue on the positive side of this line corresponding to the tearing mode. An example of the spectrum is given in Figure \ref{spectrum}. This spectrum has been produced with $S=1000$, $k=0.5$ and $N=700$.  The spectrum consists of three main branches (labelled $b_1$, $b_2$ and $c$) and two (sub)branches connecting $b_1$ and $b_2$ to $c$, labelled $s_1$ and $s_2$. This branching structure is qualitatively similar to that found by \cite{riedel86} by means of WKB analysis and also matches other numerical studies \citep{dmac17,adv}. The spectrum is also symmetric about $\Imag(\sigma)=0$. It can easily be demonstrated that equation (\ref{eig_new}) posseses the symmetry
\be\label{sym}
\sigma\rightarrow\sigma^*,\quad u\rightarrow-u^*, \quad b\rightarrow b^*,
\en
for the background equilibria we have selected. The asterisk denotes the complex-conjugate.

\begin{figure}

	\centering
	
		{\includegraphics[scale=0.6,keepaspectratio]{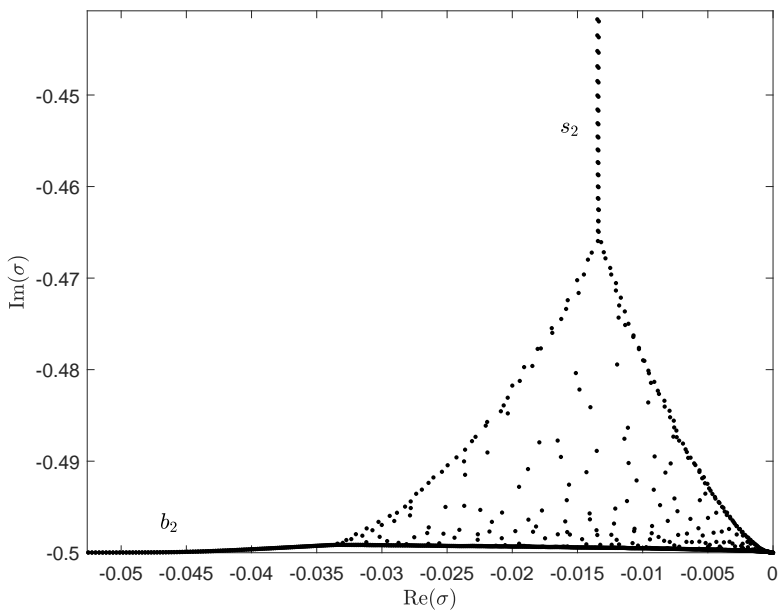}}
	\caption{Eigenvalue spectrum for $S=10^6$ and $k=0.5$ at the intersection of branches $b_2$ and $s_2$.}
\label{spectrum2}
\end{figure}

The general branching structure outlined above is maintained as $S$ is increased, however, the intersections of the $b$ and $s$ branches become more compressed. To illustrate this, Figure \ref{spectrum2} displays the part of the spectrum, for the case $S=10^6$, $k=0.5$ and $N=2000$, where the branches $b_2$ and $s_2$ meet. By the symmetry of (\ref{sym}), there is an equivalent structure at the intersection of $b_1$ and $s_1$.

The $b_2$ branch has been pushed closer to the $\Real(\sigma)=0$ line and the eigenvalues have become densely packed into a small triangular region. This mirrors the behaviour of the spectrum of the Orr-Sommerfield operator at the intersection of its eigenvalue branches for large $Re$ \citep[e.g.][]{sh01}. Eigenvalues in such intersection regions are highly sensitive and full numerical convergence is difficult to achieve \citep{kerner98}. However, it is this sensitivity that makes these regions important for transient growth, as will be demonstrated later. Despite the numerical difficulties associated with the calculation of these spectra, our transient growth calculations, which depend on the spectra, are converged and will be shown to follow a clear scaling law (see \cite{hanifi96} for a similar situation).

\subsection{Pseudospectra}
The individual eigenvalues and eigenvectors only describe the behaviour of a linear instability on a large time scale. They give no information about transient growth that can occur much sooner. One mathematical structure which can provide information about transient growth is a generlization of the spectrum known as the pseudospecturm \citep{t05}. 
\begin{definition}
Let $\am=\cm^{-1}\cl$ and $\epsilon>0$ be arbitrary. The $\epsilon$-pseudospectrum $\se(\am)$ of $\am$ is the set of $z\in\mathbb{C}$ such that
\be
\|(z\mathsfbi{I}-\mathsfbi{A})^{-1}\|>\epsilon^{-1},
\en
where $\mathsfbi{I}$ is the identity matrix. 
\end{definition}
Note that in this definition we could replace $\am$ by $\bm$, as defined in Section \ref{quad}. All numerical calculations involving norms will use $\bm$ in this article due to the spectral discretization. Note also that pseudospectra are not related to the pseudospectral discretization of the differential equations described in the previous section, both topics just share the same name.

\begin{figure}

	\centering
	
		{\includegraphics[scale=0.6,keepaspectratio]{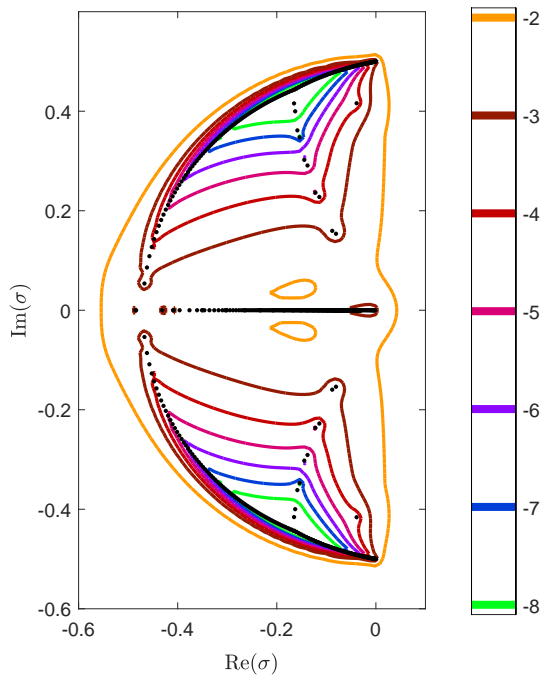}}

	\caption{The eigenvalue spectrum and pseudospectra for the projection covering $-0.5<\Real(\sigma)<0$. Eigenvalues are shown as solid dots. The boundaries of $\epsilon$-pseudospectra are for $\epsilon= 10^{-8},\dots,10^{-2}$.}
\label{pseudospectrum}
\end{figure}

To quote the monograph on pseudospectra, \cite{t05} (to which the reader is directed for a comprehensive account of the subject),  ``the $\epsilon$-pseudopectrum is the open subset of the complex plane bounded by the $\epsilon^{-1}$ level curve of the norm of the resolvent''. For non-normal matrices, these level curves can extend $O(1)$ distances from the position of the eigenvalues. For normal matrices, the curves form $O(\epsilon)$ balls around the eigenvalues. The $\epsilon$-pseudospectra for the projection of $\mathsfbi{A}$, where the eigenvalues satisfying  $-0.5<\Real(\sigma)<0$ are kept, are shown in Figure \ref{pseudospectrum} for a range of $\epsilon$. This figure has been produced using EigTool \citep{wright02}.

The extension of the level curves far from the position of the eigenvalues, particularly at the intersection points of the $b$ and $s$ branches, is a clear sign of the non-normality of $\mathsfbi{A}$. Further, a level curve is displayed crossing $\Real(\sigma)=0$ into the positive half plane. This fact gives important information on the behaviour of transient growth. To see why, first note that the formal solution of the discretized version of equation (\ref{ivp}) can be written as
\be\label{formal}
\vv(t) = \exp(t\mathsfbi{A})\vv(0).
\en

In normal mode analysis, the growth rate of the linear onset of an instability is given by the rightmost eigenvalue,
\be
\alpha(\mathsfbi{A}) = \sup_{z\in\sigma(\mathsfbi{\mathsfbi{A}})}\Real(z).
\en 
The quantity $\alpha(\mathsfbi{A})$ is also known as the spectral abscissa of $\mathsfbi{A}$. An analogous definition for pseudospectra is
\be
\alpha_{\epsilon}(\mathsfbi{A}) = \sup_{z\in\se(\mathsfbi{A})}\Real(z).
\en
The envelope of transient growth is given by $\|\exp(t\mathsfbi{A})\|$ for $t\ge 0$. {This envelope gives the maximum possible transient growth that can be achieved at a time $t$ optimized over all normalized initial conditions \citep{sh01}. This fact can found easily from equation (\ref{formal}), i.e. 
\be
\sup_{\vv(0)}\frac{\|\vv(t)\|_2^2}{\|\vv(0)\|_2^2} = \sup_{\vv(0)}\frac{\|\exp(t\mathsfbi{A})\vv(0)\|_2^2}{\|\vv(0)\|_2^2}=\|\exp(t\mathsfbi{A})\|_2^2.
\en
} A simple and practical lower bound on the envelope height is given by  
\be\label{psbound}
\sup_{t\ge 0}\|\exp(t\mathsfbi{A})\|\ge\frac{\alpha_{\epsilon}(\mathsfbi{A})}{\epsilon}\quad \forall\epsilon>0.
\en
This result is related to the Kreiss matrix theorem \citep{t05}. 

For our choice of $\mathsfbi{A}$, $\alpha(\am)<0$ describes only an asymptotically decaying solution. Since the pseudospectra in Figure \ref{pseudospectrum} pass beyond $\Real(\sigma)=0$, the bound in (\ref{psbound}) states that transient growth can be expected.   Looking at the level curve for $\epsilon=10^{-2}$ in Figure \ref{pseudospectrum}, the maximum value of $\Real(z)\approx0.023$. Hence, it follows from (\ref{psbound}) that $\sup_{t\ge 0}\|\exp(t\am)\|\gtrsim2.23$. Later we will demonstrate that the maximum of transient growth follows a precise scaling as a function of $S$.
\subsection{Transient growth dependence on the spectrum}\label{tdep}
\begin{figure}

	\centering
		{\includegraphics[scale=0.7,keepaspectratio]{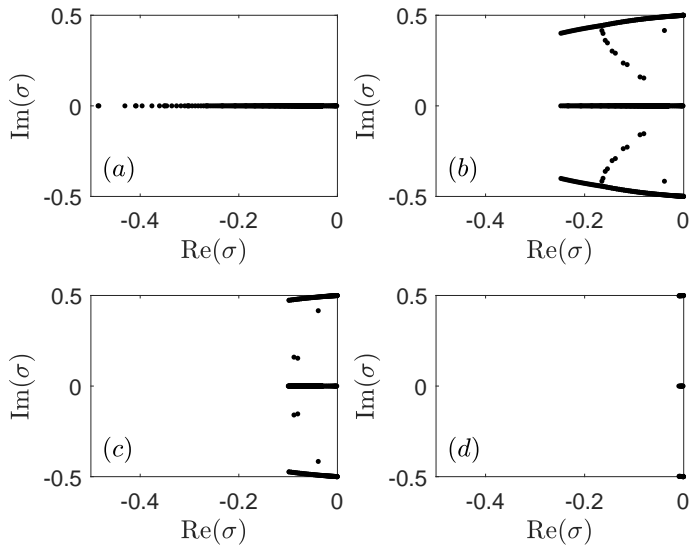}}

	\caption{The eigenvalue spectra of projections of $\mathsfbi{A}$ focussing on different parts of $\mathbb{C}$. Details are given in the main text.}
\label{Tbreak}
\end{figure}

In order to undertand how the transient growth depends on different parts of the spectrum we will consider four different projections of $\mathsfbi{A}$  (using the parameters $S=10^3$, $k=0.5$ and $N=700$) that focus on different parts of $\mathbb{C}$, i.e. on different groups of eigenvalues. Figure \ref{Tbreak} displays the eigenvalue spectra of the four cases that we consider.
Case (a) contains only the $c$ branch with eigenvalues satisfying $-0.5<\Real(\sigma)<0$. Case (b) includes parts  of all three main branches ($b_1$, $b_2$ and $c$) and branches $s_1$ and $s_2$.  These eigenvalues are in the range $-0.25<\Real(\sigma)<0$. Case (c) omits the branch connections of $b_1$ with $s_1$ and $b_2$ with $s_2$. These eigenvalues are in the range $-0.1<\Real(\sigma)<0$. Case (d) considers a small selection of eigenvalues in the range $-0.01<\Real(\sigma)<0$, which lie on the three main branches.

\begin{figure}

	\centering
		
		{\includegraphics[scale=0.7,keepaspectratio]{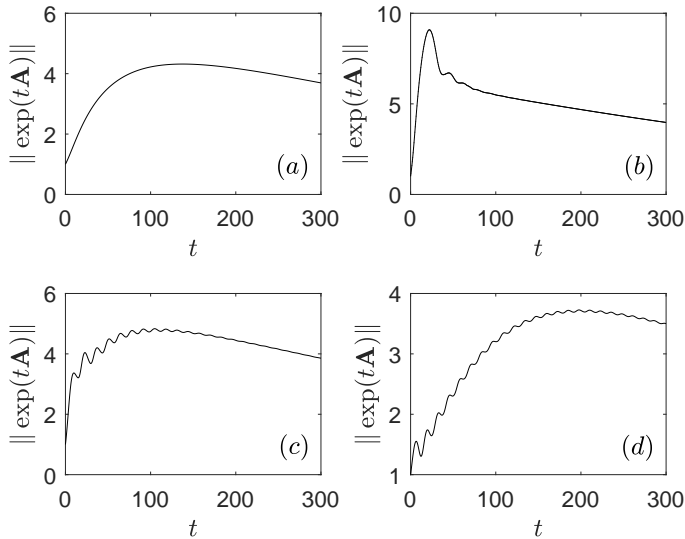}}

	\caption{The transient growth envelopes for the four projections whose spectra are displayed in Figure \ref{Tbreak}.}
\label{Ebreak}
\end{figure}
Looking at the overall behaviour of the transient growth displayed in Figure \ref{Ebreak}, cases (a), (c) and (d) are similar in that all reach maximum growth at $t\approx100-200$ and have similar maxima in the range 3.5-5. The main visual difference between these cases is that the transient growth in case (a) is smooth, whereas those in cases (c) and (d) possess many oscillations. This difference is down to only the central branch of eigenvalues being included in (a), whereas portions of the three main branches are included in (c) and (d). Case (c) also has a faster initial growth rate compared to cases (a) and (d) and attains the highest transient growth out of the three cases. The spectrum of case (c) contains the upper parts of the $s$ branches but not the conneciton points with the $b$ branches. 

 Case (b) is strikingly different to the rest. Its transient growth exhibits a sharp rise to a maximum that is approximately double that of the other cases. The eigenvalues considered for case (b) include all the branches and the intersection points. The transient growth curve in Figure \ref{Ebreak} (b) and the behaviour of the $\epsilon$-pseudospectra in Figure \ref{pseudospectrum} indicate that the intersection points of the $b$ and $s$ branches in the spectrum are important for strong transient growth.

\subsection{The relationship of physical quantities to the spectrum and transient growth}
The previous subsection demonstrated that {including} the branching structure of the eigenvalue spectrum {in the projection} is important for fast and strong transient growth. We will now focus on how the branches depend on physical quantities, namely forces and energy balance. 

Let us first consider forces in the momentum balance equation (\ref{mhd1}). In the present setup of the TI with no equilibrium flow, the viscosity plays little role. This can be seen by comparing Figure \ref{spectrum} with Figure 1(a) of \cite{dmac17}, which plots the spectrum for the same parameters but for the inviscid MHD equations. Note that in \cite{dmac17}, time dependence is based on $\exp(-i\sigma t)$ rather than $\exp(\sigma t)$, so the plot of the spectrum is rotated. 

\begin{figure}

	\centering
	
		{\includegraphics[scale=0.45,keepaspectratio]{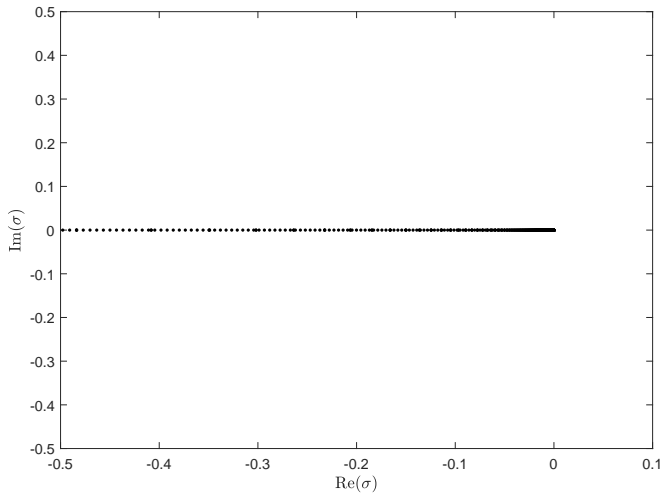}}

	\caption{The spectrum of kinematic MHD with the same parameters used for Figure \ref{spectrum}.}
\label{kinematic}
\end{figure}

In order to assess the effect of the Lorentz force, we can compare the spectrum of resistive MHD to that of kinematic MHD. Kinematic MHD is concerned with the solution of the induction equation and normally does not consider the momentum equation (the velocity field is prescribed). Here, we can solve a form of kinematic MHD where the momentum equation is solved but the Lorentz force is removed. In the linearized equations, this is equivalent to setting $L_{B_0}=0$. Figure \ref{kinematic} displays the spectrum of kinematic MHD for the same parameters used for Figure \ref{spectrum}. 

Notice that the $c$ branch is the only one remaining and that this system is asymptotically stable to linear perturbations. From the results of the previous subsection, this means that the possible transient growth is much more limited. This result makes sense physically since the energy source of the TI is the equilibrium magnetic field and the bending of field lines via the Lorentz force can lead to larger perturbations. 

Another way to consider what physical quantities play a role in trasient growth is to determine the energy balance. \cite{borba94} did this for inviscid MHD and we will extend the analysis for the present visco-resistive case. Using the notation of Section 2, the application of vector calculus leads to
\begin{eqnarray}\label{eb}
\frac{\rm d}{{\rm d}t}\frac12\int(|\uv_1|^2+|\Bv_1|^2)\,{\rm d}V&=& \int[\jv_1\cdot(\uv_0\times\Bv_1)-\jv_0\cdot(\uv_1\times\Bv_1)]\,{\rm d}V-\frac{1}{S}\int|\nabla\times\Bv_1|^2\,{\rm d}V \nonumber \\
&&-\int(\uv_1\cdot\nabla)\uv_0\cdot\uv_1\,{\rm d}V - \frac{1}{Re}\int|\nabla\times\uv_1|^2\,{\rm d}V,
\end{eqnarray}
{where $\jv_0=\nabla\times\Bv_0$ and  $\jv_1=\nabla\times\Bv_1$. The integrals associated with the diffusion terms are either zero or negative (including the negative sign)}, so these terms can only describe decay and not transient growth. The first integral on the right-hand side describes how energy can be taken from the equilibrium magnetic field {and background flow due to the effect of the Lorentz force}, leading to transient growth.  The third integral on the right-hand side describes how transient growth is possible, {without the aid of the Lorentz force}, by {the perturbed flow} extracting energy from the equilibrium flow. In the examples considered so far, {the terms involving $u_0$ are zero}. This fact, combined with the large Reynolds number we have selected ($Re=10^6$) explains the similarity of the spectra of the viscous and inviscid cases.

\subsection{Scaling of maximum transient growth}
\begin{figure}

	\centering
	
		{\includegraphics[scale=0.45,keepaspectratio]{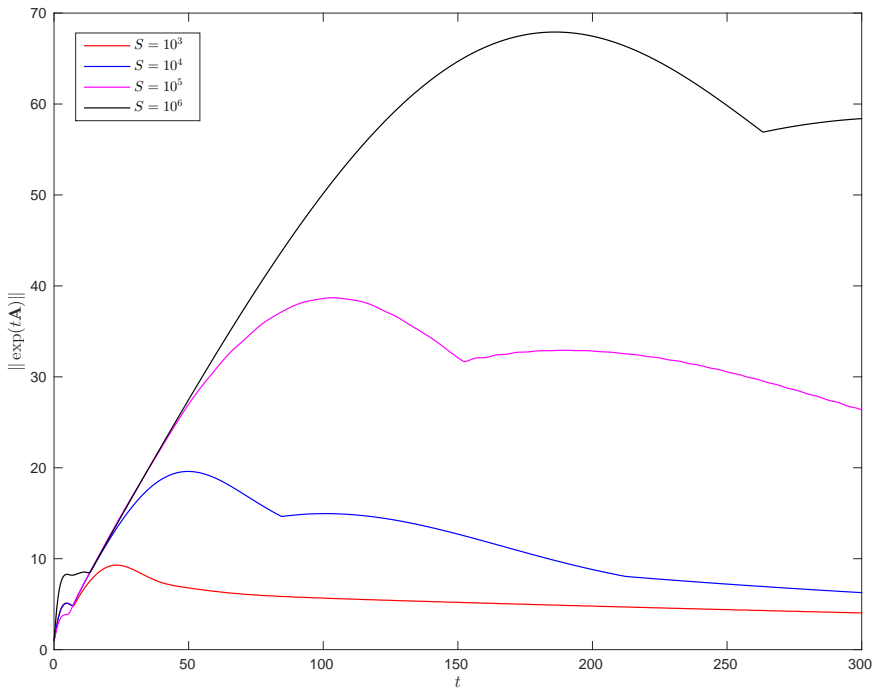}}

	\caption{The transient growth envelopes for $k=0.5$ and $S=10^3,\dots,10^6$}
\label{tcurves}
\end{figure}

We now return to the projection of $\mathsfbi{A}$ with eigenvalues in the range $-0.5<\Real(\sigma)<0$. {Since we are considering the case $k=0.5$, we take $\Re(\sigma)=-0.5$ as the `lower boundary' of our projection. Beyond this point, only the $c$ branch continues and this part of the spectrum does not contribute significantly to transient growth, as would be expected from the analysis above.} In order to determine how the transient growth scales with the Lundquist number $S$, we explicitly calculate $\|\exp(t\am)\|$ for $t\in[0,300]$ and $S=10^n$, $n=3,4,5,6$. The transient growth envelopes are shown in Figure \ref{tcurves} and are calculated with $N=2000$.

\begin{figure}

	\centering
	
		{\includegraphics[scale=0.6,keepaspectratio]{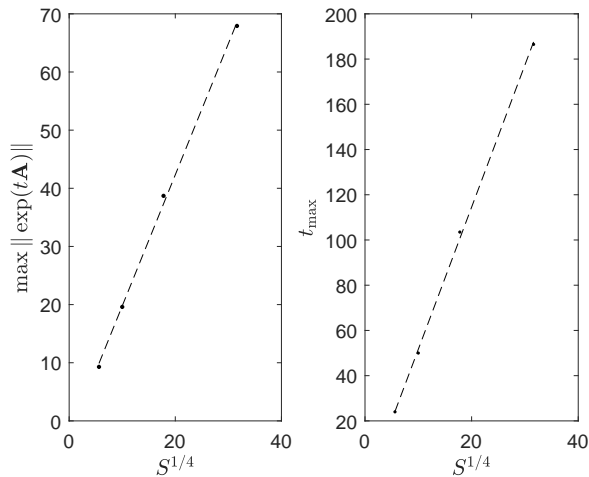}}

	\caption{The maximum transient growth, $\max\|\exp(t\am)\|$, and the time when it occurs, $t_{\max}$, as a function of $S^{1/4}$. The dashed lines are lines of best fit.}
\label{linefit}
\end{figure}

After an initial perturbation, all the curves follow the same gradient before they each turn toward their maxima. Each curve contains a `hump' where it reaches its maximum and this is due to the inclusion of the branch intersection points of $b$ and $s$ in the spectrum, as in case (b) from subsection \ref{tdep}. If we consider the maximum transient growth of each curve and the times when the maxima occur, we can find a simple scaling law. These quantities are displayed in Figure \ref{linefit}.

\begin{figure}

	\centering
	
		{\includegraphics[scale=0.7,keepaspectratio]{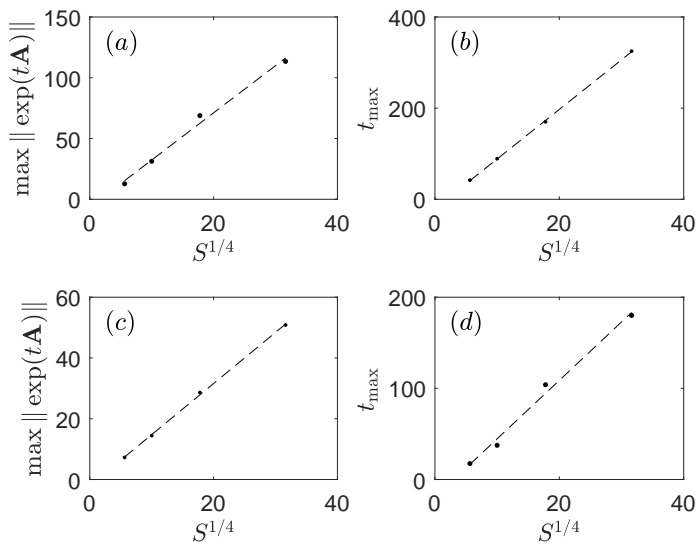}}

	\caption{The maximum transient growth, $\max\|\exp(t\am)\|$, and the time when it occurs, $t_{\max}$, as a function of $S^{1/4}$. The dashed lines are lines of best fit. (a) and (b) refer to $k=0.2$, (c) and (d) refer to $k=0.8$.}
\label{otherk}
\end{figure}

Figure \ref{linefit} demonstrates that the maximum transient growth, $\max\|\exp(t\am)\|$, and the time at which it occurs, $t_{\rm max}$, both depend linearly on $S^{1/4}$. This simple scaling relation is robust for `tearing-unstable' wave numbers, i.e. $0< k<1$. Figure \ref{otherk} demonstrates the same scaling profiles for $k=0.2$ and $k=0.8$.

{The projection we have used for the cases $k=0.2$ and $k=0.8$ is the same as that for the $k=0.5$ case. This means that for the $k=0.2$ case, more eigenvalues on the $c$ branch are used beyond the point where the $c$ and $b$ branches meet. For the $k=0.8$ case, however, the spectrum is truncated before the $b$ branches meet the $c$ branch. Despite these different truncations of the spectra, since both cases include the vital (as emphasized by the previous analysis) branch points, the optimal transient growth is insensitive to the different truncation locations and follows the same scaling law as the $k=0.5$ case.} Our numerical results suggest the scaling that a maximum transient growth of $O(S^{1/4})$, optimized using the $L_2$-norm, is possible in a time of $O(S^{1/4})$ {for} $S\gg 1$.

\subsection{Optimal initial perturbations}
\begin{figure}

	\centering
	
		{\includegraphics[scale=0.55]{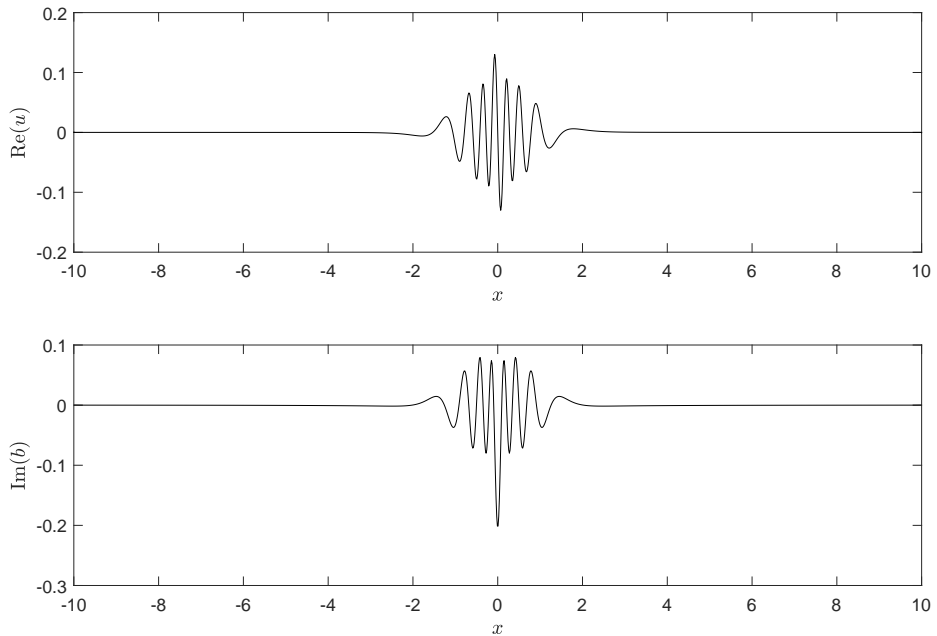}}

	\caption{The optimal initial conditions that produce the maximum transient growth at $t=50$ for $S=10^4$ and $k=0.5$.}
\label{initial}
\end{figure}

As well as determining the maximum possible transient growth at any time $t$, we can also determine the initial condition which produces that growth at time $t$. This can be achieved via a singular value decomposition (SVD),
\be
\exp(t\mathsfbi{A}) = \mathsfbi{U}\boldsymbol{\Sigma}\mathsfbi{V}^*,
\en
where $\mathsfbi{U}$ and $\mathsfbi{V}$ are unitary matrices and $\boldsymbol\Sigma$ is a matrix containing the singular values ordered by size. The asterisk denotes the complex-conjugate transpose. The first column of $\mathsfbi{V}$ corresponds to the optimal initial condition \citep{sh01,t05}. To illustrate the form of such initial conditions, Figure \ref{initial} displays the the non-zero parts of $u$ and $b$ at $t=0$ which produce the maximum transient growth at time $t=50$, for $S=10^4$ and $k=0.5$. 
The forms shown in Figure \ref{initial} have a striking resemblence to `wave packets' located at the current sheet. Such wave packet solutions can be understood with the help of pseudospectral theory. A definition of pseudospectra equivalent to Definition 1 is
\begin{definition}
$\se(\mathsfbi{A})$ is the set of $z\in\mathbb{C}$ such that
\be
\|(z\mathsfbi{I}-\mathsfbi{A})\mathbf{p}\|<\epsilon,
\en
for some vector $\mathbf{p}$ with $\|\mathbf{p}\|=1$.
\end{definition}

The vector $\mathbf{p}$ is known as a pseudomode and can be thought of as a generalization of an eigenmode in the same way that a pseudospectrum is a generalization of a spectrum. That is, a pseudomode grows algebraically and an eigenmode grows exponentially. For the onset of the TI, this algebraic growth can be much faster and greater, within a fixed time, than the exponential growth of the tearing mode. Pseudomodes have a close relationship to the WKB approximation of eigenmodes \citep{obrist10}, resulting in their wave packet form. They can also be interpreted as `noise' in the system \citep{vanneste07}. Further details of the theory of pseudomodes can be found in the monograph of \cite{t05}.

\section{Conclusions}
\subsection{Summary}
In this article we have studied the non-modal onset of the classical tearing instability for visco-resistive MHD. We have paid particular attention to the eigenvalue spectrum of the linearized MHD equations. We demonstrate that the branching structure of the spectrum, which exists in the `damped half plane' of $\mathbb{C}$, is important for transient growth. We reveal this behaviour through the calculation of pseudospectra and by finding the maximum {possible} transient growth due to subsets of the spectrum. The spectrum branches are also closely linked to the Lorentz force, which is needed for strong transient growth. The importance of the Lorentz force in driving transient growth is also found from considering the energy balance of the system. A simple scaling law is determined {for tearing-unstable wavenumbers}, revealing that the maximum {possible} transient growth, {measured} in the $L_2$-norm, can grow to $O(S^{1/4})$ in a time of $O(S^{1/4})$. Optimal initial conditions which produce the maximum transient growth are shown to take the shape of wave packets and can be interpreted as noise in the system. {Although significant transient growth is possible during the linear onset of the TI for $S\gg 1$, it will only occur if the form of the initial perturbation allows it. Both non-modal and modal growth are required to give a complete description of the linear onset of the TI.}

\subsection{Discussion}
\subsubsection{Tearing-stable cases}
\begin{figure}

	\centering
	
		{\includegraphics[scale=0.65]{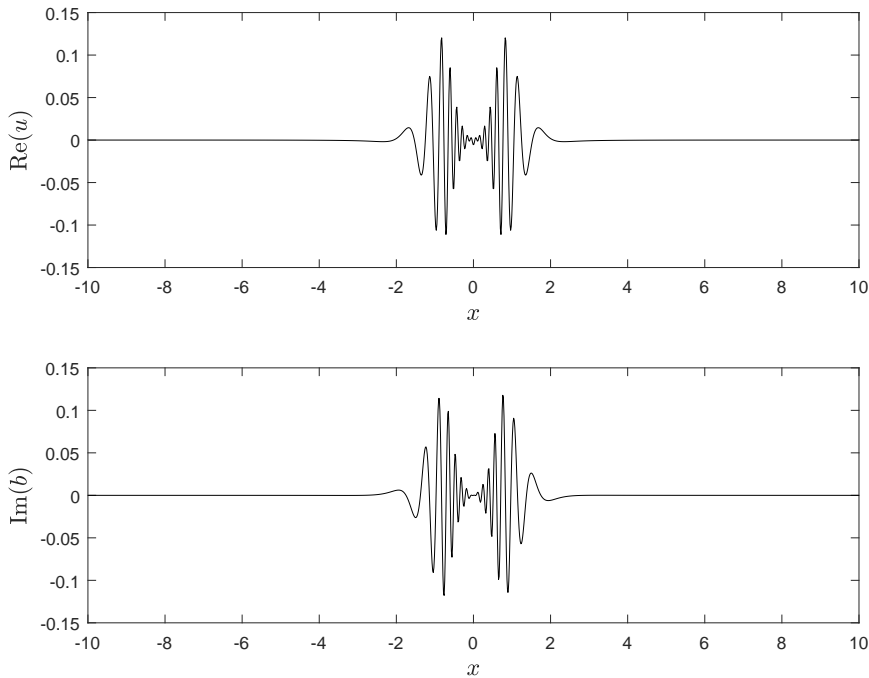}}

	\caption{The optimal initial conditions that produce the maximum transient growth at $t=50$ for $S=10^4$ and $k=1.01$.}
\label{stable_plot}
\end{figure}

{In this article we have focussed on wavenumbers for which the current sheet is unstable to the TI $(0<k<1)$. For wavenumbers $k>1$, transient growth is also possible and its maximum possible size increases with increasing $S$. There does not, however, appear to be a simple scaling law as we derived for tearing-unstable values of $k$. Figure \ref{stable_plot} displays the optimal initial conditions that produce the maximum size of transient growth at $t=50$ for $S=10^4$ and $k=1.01$ (the size of this optimal transient growth is 11.94).}

{In this example, the optimal initial conditions take the form of two wave packets on either side of the current sheet. Perturbations such as those shown in Figure \ref{stable_plot} should be used as initial conditions in non-linear MHD simulations in order to determine the non-linear consequences of linear transient growth. It may be the case that optimal initial conditions, through transient growth, can excite the tearing instability for values of $k$ which are linearly stable in normal mode analysis.}

\subsubsection{Choice of norm}
{Unlike asymptotic stability, the size of transient growth is dependent on the norm used to measure it. Optimizing with respect to different norms will give different results. Therefore, in calculations of transient growth, it is important to choose a norm with a clear physical meaning. In this article, we have focussed on the $L_2$-norm, which can be thought of as the `root mean square' of the variables and measures typical size that the variables can be amplified to. Other useful norms are the infinity norm, which measures the maximum amplitude of the perturbation and the energy norm. For a given $k$, the energy norm (disturbance kinetic plus magnetic energies)  can be written as \citep[e.g.][]{dmac17}
\be\label{energy_norm}
\|\vv\|_E^2 = \frac{1}{2k^2}\int_{-d}^{d}(|Du|^2 + k^2|u|^2 + |Db|^2 + k^2|b|^2)\,{\rm d}x = \frac{1}{2k^2}\|D\vv\|_2^2 + \frac12\|\vv\|_2^2. 
\en 
where use has been made of (\ref{mhd3})$_1$ and (\ref{mhd3})$_2$. Notice from equation (\ref{energy_norm}) that if a certain value of transient growth is found in the $L_2$-norm, this is a lower bound of the resulting energy measure. {The converse is not generally true, however, as a large measure in the energy norm need not imply a large $L_2$-norm. With regard to transient growth in the TI, however, there is still the possibility of increasing transient growth with increasing $S$, optimized with respect to the energy norm. This result was first looked at in \cite{dmac17}. Using the technique described in \cite{dmac17} we present, in Figure \ref{energy_plots}, numerical estimates of the maximum growth envelopes for different values of $S$. As in our previous analysis using the $L_2$-norm, we take $k=0.5$ and consider the contribution from asymptotically stable modes with eigenvalues in the range $-0.5<{\rm Im}(\sigma)<0$. The resolution is $N=1600$.

\begin{figure}

	\centering
	
		{\includegraphics[scale=0.65]{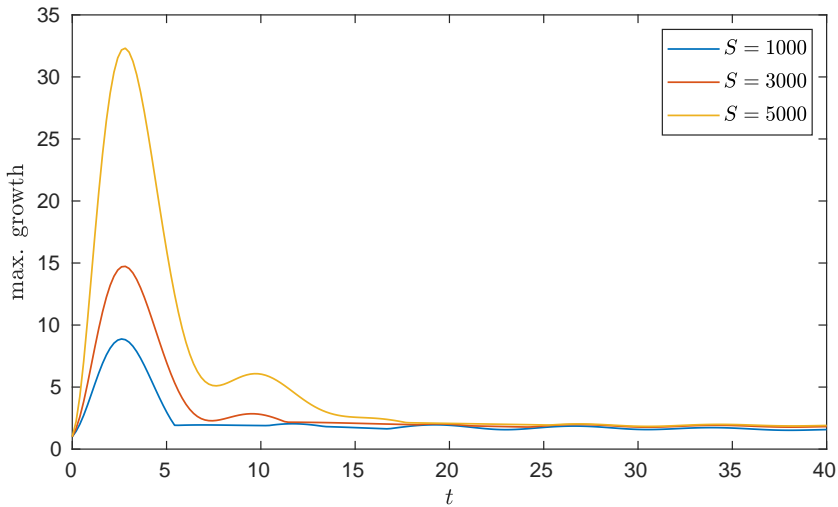}}

	\caption{The maximum energy growth envelopes for different $S$.}
\label{energy_plots}
\end{figure}
Although the shapes and maxima of the transient growth envelopes in Figure \ref{energy_plots} are different compared to those in Figure \ref{tcurves}, there is still an increasing possibility of significant transient growth as $S$ increases. Further work is required, using both the $L_2$-norm and the energy norm, to investigate transient growth at very high (astrophysical) values of $S$. }

\subsubsection{Transient growth in other TI research?}
{Simulations of the TI have generally skipped directly to the non-linear phase of the instability, e.g.  the GEM Magnetic Reconnection Challenge \citep{gem1,gem2}, or have been interpreted in terms of the tearing mode. In some high-Lundquist number simulations, such as \cite{samtaney09} with $S=10^4$, transient growth is not reported.  However, significant transient amplification will only occur if the initial condition (perturbation) is of a suitable form. It is therefore possible for transient growth not to be detected in simulations if the initial condition is not one that leads to significant transient growth. We have shown that optimal initial conditions take the form of wave packets, which can be interpreted as noise. Interestingly, recent simulations by \cite{huang17} clearly show that different levels of noise in the initial condition affect when current sheet disruption occurs {(see their Figure 12)}. It is not unreasonable to suggest that different noise patterns could lead to different transient growth, affecting when current sheet disruption takes place. A systematic study of how initial conditions, transient growth and non-linear consequences are related in MHD simulations will be carried out in future work. }

\section*{}
The author would like to thank Professor A. Valli and Doctors J. Pestana and P. Stewart for stimulating conversations on various topics related to this article. The author would also like to thank the anonymous referees for helping to improve this article.

\bibliographystyle{jpp}

\bibliography{jpp-instructions}

\begin{thebibliography}{99}
\expandafter\ifx\csname natexlab\endcsname\relax\def\natexlab#1{#1}\fi
\bibitem[Bhattacharjee et al.(2009)]{bhattacharjee09}
{\sc Bhattacharjee, A., Huang, Y.-M., Yang, H. \& Rogers, B.} 2009 Fast reconnection in high-Lundquist-number plasmas due to the plasmoid instability. {\em Phys. Plasmas \/} {\bf 16}, 112102.

\bibitem[Birn et al.(2001a)]{gem1}
{\sc Birn, J., Drake, J.~F., Shay, M.~A., Rogers, B.~N, Denton, R.~E., Hesse, M., Kuznetsova, M, Ma, Z.~W., Bhattacharjee, A., Otto, A. \& Pritchett, P.~L.} 2001a Geospace Enviromential Modelling (GEM) Magnetic Reconnection Challenge. {\em J. Geophys. Res.\/} {\bf 106}, 3715.

\bibitem[Birn et al.(2001b)]{gem2}
{\sc Birn, J. \& Hesse, M.} 2001b Geospace Environment Modelling (GEM) magnetic reconnection challenge: Resistive tearing, anisotropic pressure and Hall effects. {\em J. Geophys. Res.\/} {\bf 106}, 3737.

\bibitem[Borba et al.(1994)]{borba94}
{\sc Borba, D., Riedel, K.~S., Kerner, W., Huysmans, G.~T.~A., Ottaviani, M. \& Schmid, P.~J.} 1994 The pseudospectrum of the resistive magnetohydrodynamics operator: Resolving the resistive Alfv\'en paradox. {\em Phys. Plasmas\/} {\bf 1}, 3151.

\bibitem[Bourne(2003)]{bourne03}
{\sc Bourne, D.~P.} 2003 Hydrodynamic stability, the Chebyshev tau method and spurious eigenvalues. {\em Contin. Mech. Thermodyn.\/} {\bf 15}, 571.

\bibitem[Chen et al.(2018)]{chen18}
{\sc Chen, L., Herreman, W., Li, K., Livermore, P.~W., Luo, J.~W. \& Jackson, A.} 2018 The optimal kinematic dynamo driven by steady flow in a sphere. {\em J. Fluid Mech.\/} {\bf 839}, 1.

\bibitem[Dahlburg et al.(1983)]{dahlburg83}
{\sc Dahlburg, R.~B., Zang, T.~A., Montgomery, D. \& Hussaini, M.~Y.} 1983 Viscous, resistive magnetohydrodynamic stability computed by spectral methods. {\em Proc. Natl. Acad. Sci.\/} {\bf 80}, 5798.

\bibitem[Dahlburg(1994)]{dahlburg94}
{\sc Dahlburg, R.~B.} 1994 On the ideal initial value problem for the neutral sheet. {\em Phys. Plasmas\/} {\bf 1}, 3053.


\bibitem[Farrell \& Ioannou(1999a)]{farrell991}
{\sc Farrell, B.~F. \& Ioannou, P.~J.} 1999a Optimal exitation of magnetic fields. {\em Astrophys. J.\/} {\bf 522}, 1079.

\bibitem[Farrell \& Ioannou(1999b)]{farrell992}
{\sc Farrell, B.~F. \& Ioannou, P.~J.} 1999b Stochastic dynamics of field generation in conducting fluids. {\em Astrophys. J.\/} {\bf 522}, 1088.

\bibitem[Furth et al.(1963)]{furth63}
{\sc Furth, H.~P., Kileen, J. \& Rosenbluth, M.~N.} 1963 Finite-Resitivity Instabilities of a Sheet Pinch. {\em Phys. Fluids\/} {\bf 6}, 459.

\bibitem[Golub \& Van Loan(1996)]{golub96}
{\sc Golub, G.~H. \& Van Loan, C.~F.} 1996 Matrix Computations, 3rd ed. John Hopkins University Press.

\bibitem[Goedbloed et al.(2010)]{adv}
{\sc Goedbloed, J.~P., Keppens, R. \& Poedts, S.} 2010 Advanced Magnetohydrodynamics. Cambridge University Press. 

\bibitem[Hanifi et al.(1996)]{hanifi96}
{\sc Hanifi, A., Schmid, P.~J. \& Henningson, D.~S.} 1996 Transient growth in compressible boundary layer flow. {\em Phys. Fluids\/} {\bf 8}, 826.


\bibitem[Harris(1962)]{harris62}
{\sc Harris, E.~G.} 1962 On a plasma sheath separating regions of oppositely directed magnetic field. {\em Nuovo Cimento\/} {\bf 23}, 115.

\bibitem[Huang et al.(2017)]{huang17}
{\sc Huang, Y.-M. Comisso, L. \& Bhattacharjee, A.} 2017 Plasmoid Instability in Evolving Current Sheets and Onset of Fast Reconnection. {\em Astrophys. J.\/} {\bf 84}, 75.

\bibitem[Kerner(1998)]{kerner98}
{\sc Kerner, W.} 1998 Large-scale complex eigenvalue problems. {\em J. Comp. Phys.\/} {\bf 85}, 1. 

\bibitem[Landremann et al.(2015)]{landremann15}
{\sc Landremann, M., Plunk, G.~G. \& Dorland, W.} 2015 Generalized universal instability: transient linear amplification and subcritical turbulence. {\em J. Plasma Phys.\/} {\bf 81}, 905810501.

\bibitem[Livermore \& Jackson(2006)]{livermore06}
{\sc Livermore, P.~W. \& Jackson, A.} 2006 Transient magnetic energy growth in spherical stationary flows. {\em Proc. R. Soc. A\/} {\bf 462}, 2457.

\bibitem[Loureiro et al.(2007)]{loureiro07}
{\sc Loureiro, N.~F., Schekochihin, A.~A. \& Cowley, S.~C.} 2007 Instability of current sheets and formation of plasmoid chains. {\em Phys. Plasmas\/} {\bf 14}, 100703. 

\bibitem[MacTaggart \& Stewart(2017)]{dmac17}
{\sc MacTaggart, D. \& Stewart, P.} 2017 Optimal Energy Growth in Current Sheets. {\em Solar Phys.\/} {\bf 292}, 148.

\bibitem[Newton et al.(2010)]{newton10}
{\sc Newton, S.~L., Cowley, S.~C. \& Loureiro, N.~F.} 2010 Understanding the effect of sheared flow on microinstabilities. {\em Plasma Phys. Control. Fusion\/} {\bf 52}, 125001.

\bibitem[Obrist \& Schmid(2010)]{obrist10}
{\sc Obrist, D. \& Schmid, P.~J.} 2010 Algebraically decaying modes and wave packet pseudo-modes in swept Hiemenz flow. {\em J. Fluid Mech.\/} {\bf 643}, 309.

\bibitem[Pucci \& Velli(2014)]{pucci14}
{\sc Pucci, F. \& Velli, M.} 2014 Reconnection of quasi-singular current sheets: the ``ideal'' tearing mode. {\em Astrophys. J.\/} {\bf 780}, L19.

\bibitem[Reddy et al.(1993)]{reddy93}
{\sc Reddy, S.~C., Schmid, P.~J. \& Henningson, D.~S.} 1993 Pseudospectra of the Orr-Sommerfield Operator. {\em SIAM J. App. Math.\/} {\bf 53}, 15.

\bibitem[Riedel(1986)]{riedel86}
{\sc Riedel, K.~S.} 1986 The spectrum of resistive viscous magnetohydrodynamics. {\em Phys. Fluids\/} {\bf 29}, 1093.

\bibitem[Samtaney et al.(2009)]{samtaney09}
{\sc Samtaney, R., Loureiro, N.~F., Uzdensky, D.~A., Schekochihin, A.~A. \& Cowley, S.~C.} 2009 Formation of plasmoid chains in magnetic reconnection. {\em Phys. Rev. Lett.\/} {\bf 103}, 105004.

\bibitem[Schmid \& Henningson(2001)]{sh01}
{\sc Schmid, P.~J. \& Henningson, D.~S.} 2001 Stability and Transition in Shear Flows. Springer, New York.

\bibitem[Squire \& Bhattacharjee(2014a)]{squire14a}
{\sc Squire, J. \& Bhattacharjee, A.} 2014a Nonmodal Growth of the Magnetorotational Instability. {\em Phys. Rev. Lett.\/} {\bf 113}, 025006.

\bibitem[Squire \& Bhattacharjee(2014b)]{squire14b}
{\sc Squire, J. \& Bhattacharjee, A.} 2014b Magnetorotational instability: nonmodal growth and the relationship of global modes to the shearing box. {\em Astrophys. J.\/} {\bf 797}, 67.

\bibitem[Stewart et al.(2009)]{stewart09}
{\sc Stewart, P.~S., Waters, S.~L., Billingham, J., Jensen, O.} 2009 Spatially localised growth within global instabilities of flexible channel flow. In {\em Proceedings of the Seventh IUTAM Symposium on Laminar-Turbulent Transition\/}, IUTAM bookseries ({\bf18}), 397. 

\bibitem[Tenerani et al.(2016)]{tenerani16}
{\sc Tenerani, A., Velli, M, Pucci, F. Landi, S. \& Rappazzo, A.~F.} 2016 `Ideally' unstable current sheets and the triggering of fast magnetic reconnection. {\em J. Plasma Phys.\/} {\bf 82}, 535820501.

\bibitem[Trefethen(1999)]{t99}
{\sc Trefethen, L.~N.} 1999 Computation of pseudospectra. {\em Acta Numer.\/} {\bf 8}, 247.

\bibitem[Trefethen(2000)]{t00}
{\sc Trefethen, L.~N.} 2000 Spectral Methods in MATLAB. SIAM, Philadelphia.

\bibitem[Trefethen \& Embree(2005)]{t05}
{\sc Trefethen, L.~N. \& Embree, M.} 2005 Spectra and pseudospectra: the behaviour of non-normal matrices and operators. Princeton University Press.

\bibitem[van Wyk et al.(2016)]{vw16}
{\sc van Wyk, F., Highcock, E.~G., Schekochihin, A.~A. Roach, C.~M., Field, A.~R. \& Dorland, W.} 2016 Transition to subcritical turbulence in a tokamak plasma. {\em J. Plasma Phys.\/} {\bf 82}, 905820609. 

\bibitem[Uzdensky \& Loureiro(2016)]{uzdensky16}
{\sc Uzdensky, D.~A. \& Loureiro, N.~F.} 2016 Magnetic reconnection onset via disruption of a forming current sheet by the tearing instability {\em Phys. Rev. Lett. \/} {\bf 116}, 105003. 

\bibitem[Vanneste \& Byatt-Smith(2007)]{vanneste07}
{\sc Vanneste, J. \& Byatt-Smith, J.~G.} 2007 Fast scalar decay in a shear flow: modes and pseudomodes. {\em J. Fluid Mech.\/} {\bf 572}, 219. 


\bibitem[Wright(2002)]{wright02}
{\sc Wright, T.~G.} 2002 EigTool. \texttt{http://www.comlab.ox.ac.uk/pseudospectra/eigtool/}.

\end{thebibliography}

\end{document}